\documentclass[draftcls,a4paper,onecolumn,12pt]{IEEEtran}
\usepackage{graphicx}
\usepackage{amsmath}
\usepackage{amssymb}
\usepackage{algorithmic}
\usepackage{algorithm}
\usepackage{hyperref}
\RequirePackage[normalem]{ulem}

\newif\iffigures

\figurestrue

\graphicspath{{./}{/home/jungp/TeX/Figures/}}

\usepackage{graphics,graphicx,color}

\newcommand{\Trace}[1]{\,\mathbf{Tr}{#1}}
\newcommand{\setZ}{{\mathbb{Z}}}
\newcommand{\Ztwo}{\setZ^2}
\newcommand{\Gabor}{{\mathcal{G}}}

\newcommand{\setblochrankone}{\mathcal{B}^1}
\newcommand{\setDO}{\mathfrak{S}}
\newcommand{\setrankone}{Z}
\newcommand{\settraceideal}{\setDO}
\newcommand{\schattenclass}{\mathcal{T}}
\newcommand{\settraceclass}{\schattenclass_1}
\newcommand{\without}{\setminus}
\newcommand{\order}{o}

\newcommand{\defeq}{\overset{\text{def}}{=}}

\newcommand{\Bx}{\boldsymbol{x}}

\newcommand{\Bn}{\boldsymbol{n}}
\newcommand{\HH}{{\mathcal{H}}}
\newcommand{\BH}{\boldsymbol{\HH}}

\newcommand{\sigmaN}{{\sigma^2}}

\newcommand{\BHSpread}{\boldsymbol{\Sigma}}

\newcommand{\BHScat}{\boldsymbol{C}}

\newcommand{\major}{\succ}
\newcommand{\minor}{\prec}

\newcommand{\diag}{\text{diag}}

\newcommand{\Leb}[1]{\mathcal{L}_{#1}}
\newcommand{\Ltwo}{\Leb{2}}

\newcommand{\Indexset}{{\mathcal{I}}}

\newcommand{\Id}{{\mathbb{I}}}

\newcommand{\Amb}{{\mathbf{A}}}

\newcommand{\SINR}{\text{\small{\rm{SINR}}}}

\newcommand{\EX}[1]{{\mathbf{E}}\{#1\}}
\newcommand{\Ex}[2]{{\mathbf{E}}_{#1}\{#2\}}

\newcommand{\Real}[1]{{\text{Re}}\{#1\}}

\newcommand{\Shift}{{\boldsymbol{S}}}

\newcommand{\Ind}[2]{\chi_{#1}(#2)}

\newcommand{\fourier}[1]{\hat{#1}}

\newcommand{\Reals}{\mathbb{R}}
\newcommand{\RealsPlus}{\mathbb{R}_{+}}
\newcommand{\Complexes}{\mathbb{C}}

\newcommand{\taumax}{{\tau_d}}

\renewcommand{\taumax}{{\tau_d}}

\begin{document}
\title{The WSSUS Pulse Design Problem in Multicarrier Transmission}
\author{Peter Jung and Gerhard Wunder \\
  Fraunhofer German-Sino Lab for Mobile Communications - MCI \\[.1em]
  \small{\{jung,wunder\}@hhi.fraunhofer.de}}
\maketitle
\begin{abstract}
   Optimal link adaption to the scattering function of wide sense stationary uncorrelated scattering (WSSUS)
   mobile communication channels is still an unsolved problem despite its importance 
   for next-generation system design.
   In multicarrier transmission such link adaption is performed by pulse
   shaping, i.e. by properly adjusting the transmit and receive filters.
   For example pulse shaped Offset--QAM systems have been recently
   shown to have superior performance over standard cyclic prefix OFDM
   (while operating at higher spectral efficiency).
   In this paper we establish a general mathematical framework for joint
   transmitter and
   receiver pulse shape optimization for so-called Weyl--Heisenberg or Gabor
   signaling with respect to
   the scattering function of the WSSUS channel. In our framework the pulse shape
   optimization problem is translated
   to an optimization problem over trace class operators which in turn
   is related to fidelity optimization in quantum information processing.
   By convexity relaxation the problem is shown to be equivalent to a
   \emph{convex constraint quasi-convex maximization problem} thereby
   revealing the non-convex nature of the overall WSSUS pulse
   design problem. We present several iterative algorithms for optimization 
   providing applicable results even for large--scale problem constellations.
   We show that with transmitter-side knowledge of the channel statistics
   a gain of $3 - 6$dB in $\SINR$ can be expected.
\end{abstract}
\section{Introduction}
It is well known that channel information at the transmitter increases link capacity. 
However, since future mobile communication is expected to operate in fast varying channels,
it is not realistic to assume perfect channel knowledge. On the other hand statistical 
information can be used which does not change in the rapid manner as the channel itself.
In multicarrier communications this can be employed for the design of transmitter
and receiver pulse shapes. Unfortunately the problem of optimal signaling in this context is
still an unsolved problem.
Orthogonal Frequency Division Multiplexing (OFDM) has the capability to resolve the 
inherent structure of time-invariant (or slowly fading) channels, i.e. converts the frequency-selective
channel into flat fading channels on the corresponding subcarriers.
From mathematical point of view the joint transmitter and receiver signaling (that includes an 
appropriate cyclic prefix) diagonalizes a complete class of linear time-invariant channels.
Apart from the cyclic prefix, which implies bandwidth and power efficiency loss
the classical OFDM approach seems to be an efficient setup for the time-invariant case. 
But this does not hold anymore if we consider additional time-variant distortions
caused by the mobile
channel or non-ideal radio frontends \cite{jung:ieeecom:timevariant}.
The conventional OFDM scheme can be extended in several ways 
to match the requirements for more mobility and increased bandwidth efficiency.
In particular an approach based on Offset--QAM (OQAM) \cite{chang:oqam} in conjunction with 
Gaussian-like pulse shapes (OQAM/IOTA) \cite{lefloch:cofdm}, 
also incorporated in the 3GPP OFDM study item \cite{lacroix:vtc01,jung:inowo2004}, 
reflects a promising new direction. 
Due to the enhancement of the physical layer an improvement of the overall system 
performance is expected while the air interface is still very similar to OFDM.
Thus, it makes sense to consider a more general signaling, 
namely Weyl-Heisenberg signaling, and assess the problem of optimal pulse shapes for a given
second order statistics of the time-variant scattering environment.

The paper is organized as follows. In the first part we formulate, what is mainly known under
the name Weyl--Heisenberg (or Gabor) signaling. Then we will review in this context 
cyclic prefix based OFDM (cp--OFDM) and pulse shaped OQAM. In the second part of 
the paper we present the principles of WSSUS pulse adaption for Weyl--Heisenberg signaling. 
We will establish 
the main optimization functional and show several design strategies for its maximization. 
Then we give a more abstract formulation and identify
pulse optimization as convex--constraint (quasi-) convex maximization problems.
In the third part 
we then explicitely work out the optimization strategies and algorithms. 
Finally the performance of the iterative algorithms is evaluated.

\section{System Model}
Conventional OFDM and pulse shaped OQAM can be jointly formulated within the concept of 
generalized multicarrier schemes, which means that
some kind of time--frequency multiplexing will be performed.
To avoid cumbersome notation we will adopt a two--dimensional index notation
$n=(n_1,n_2)\in\setZ^2$ for time--frequency slots $n$.
In our framework the baseband transmit signal  is
\begin{equation}
   \begin{aligned}
      s(t)
      =\sum_{n\in\Indexset}x_n\gamma_n(t)
      =\sum_{n\in\Indexset}x_n(\Shift_{\Lambda n}\,\gamma)(t)
   \end{aligned}
   \label{equ:txsignal}
\end{equation}
where $(\Shift_{\mu}\,\gamma)(t)\defeq e^{i2\pi \mu_2t}\gamma(t-\mu_1)$ ($i$ is the imaginary unit
and $\mu=(\mu_1,\mu_2)$) is a time-frequency shifted 
version of the transmit pulse $\gamma$, i.e. $\gamma_n\defeq\Shift_{\Lambda n}\,\gamma$ is shifted according to 
a lattice $\Lambda\Ztwo$ ($\Lambda$ denotes its $2\times 2$ real generator matrix).
The indices $n=(n_1,n_2)$  range over 
the doubly-countable set $\Indexset\subset\Ztwo$, referring to the data burst to be transmitted. 
Due to this lattice structure in the time-frequency plane (or phase space), 
this setup is sometimes called Gabor signaling. 
Moreover, because the time-frequency shift operators (or phase space displacement operators)
$\Shift_\mu$ are unitary representations of the Weyl-Heisenberg group (see for example
\cite{folland:harmonics:phasespace,miller:topicsharmonic}) on 
$\Ltwo(\mathbb{R})$ this also known as Weyl-Heisenberg signaling. 
In practice $\Lambda$ is often restricted to be diagonal, i.e. $\Lambda=\diag(T,F)$.
However, Gabor based multicarrier transmission can generalized to
other lattices as well \cite{strohmer:lofdm2}.
The time-frequency sampling density is related to the 
bandwidth efficiency (in complex symbols) of the signaling, i.e. $\epsilon\defeq|\det\Lambda^{-1}|$, which
gives $\epsilon=(TF)^{-1}$ for $\Lambda=\diag(T,F)$.

The coefficients $x_n$ are the complex 
data symbols at time instant $n_1$ and subcarrier index $n_2$ with the property
$\EX{\Bx\Bx^*}=\Id$ (from now on $\,\bar{\cdot}$ always denotes complex conjugate and
$\cdot^*$ means conjugate transpose), where $\Bx=(\dots,x_n,\dots)^T$.
We will denote the linear time-variant channel by $\BH$
and the additive white Gaussian noise process (AWGN) by $n(t)$.
The received signal is then
\begin{equation}
   \begin{aligned}
      r(t)=(\BH s)(t)+n(t)
      =\int_{\Reals^2} \BHSpread(\mu)(\Shift_\mu s)(t)d\mu + n(t)
   \end{aligned}
\end{equation}
with $\BHSpread:\Reals^2\rightarrow\Complexes$ 
being a realization of the (causal) channels spreading
function with finite support. 
We used here the notion of the {\it wide-sense stationary uncorrelated
  scattering} (WSSUS) channel \cite{bello:wssus} and its decomposition into time-frequency shifts. In the WSSUS assumption the
channel is characterized by the second order statistics of $\BHSpread(\cdot)$, i.e. the
scattering function $\BHScat:\Reals^2\rightarrow\RealsPlus$
\begin{equation}
   \begin{aligned}
      \EX{\BHSpread(\mu)
        \overline{\BHSpread(\mu')}}=\BHScat(\mu)\delta(\mu-\mu')
   \end{aligned}
   \label{eq:wssus:assumptions}
\end{equation}
Moreover we assume $\EX{\BHSpread(\mu)}=0$. Without loss of generality we use
$\lVert\BHScat\rVert_1=1$, which means that the channel has no overall path loss.
To obtain the data symbol $\tilde{x}_{m}$ the receiver projects 
on $g_m\defeq\Shift_{\Lambda m} g$ with $m\in\Indexset$, i.e.
\begin{equation}
   \begin{aligned}
      \tilde{x}_{m}
      &=\langle g_m,r\rangle
      =\langle \Shift_{\Lambda m} g,r\rangle
      =\int \,e^{-i2\pi (\Lambda m)_2t}\overline{g(t-(\Lambda m)_1)}\,r(t)dt \\
   \end{aligned}
   \label{equ:receivedsymbol}
\end{equation}
By introducing the elements 
\begin{equation}
   \begin{aligned}
      H_{m,n}&\defeq\langle g_m,\BH\gamma_n\rangle=
      \int_{\Reals^2}\Sigma(\mu) \langle g_m,\Shift_\mu\gamma_n\rangle d\mu\\
   \end{aligned}
\end{equation}
of the channel matrix $H\in\mathbb{C}^{\Indexset\times\Indexset}$, 
the multicarrier transmission can be formulated as the linear equation
\mbox{$\tilde{\Bx}=H\Bx+\tilde{\Bn}$},
where $\tilde{\Bn}=(\dots,\langle g_m,n\rangle,\dots)^T$ is the vector of the projected noise
having variance $\sigmaN:=\Ex{n}{|\langle g_m,n\rangle|^2}$ per component.
If we assume that the receiver has perfect channel knowledge
(given by $\BHSpread$) ''one--tap''
(zero forcing) equalization would be of the form
$\tilde{x}^{\text{eq}}_m=H_{m,m}^{-1} \tilde{x}_{m}$ 
(or alternatively MMSE equalization if $\sigmaN$ is known), where
\begin{equation}
   \begin{aligned}
      H_{m,m}
      &=\int_{\Reals^2}\Sigma(\mu)e^{i2\pi(\mu_1(\Lambda m)_2-\mu_2(\Lambda m)_1)}\langle g,\Shift_\mu\gamma\rangle d\mu
      =\int_{\Reals^2}\Sigma(\mu)e^{i2\pi(\mu_1(\lambda m)_2-\mu_2(\lambda m)_1)}\Amb_{g\gamma}(\mu)d\mu
   \end{aligned}
\end{equation}
Here $\Amb_{g\gamma}(\mu)\defeq\langle g,\Shift_\mu\gamma\rangle$ 
is the well known cross ambiguity function of $g$ and $\gamma$.

We adopt the following $\ell_2$--normalization of the pulses. The normalization of $g$ will have no effect on
the later used system performance measures. The normalization of $\gamma$ is typically
determined by some transmit power constraint and will scale later only the noise variance
$\sigmaN\rightarrow\sigmaN/\lVert\gamma\rVert^2_2$. Thus
we assume $g$ and $\gamma$ to be normalized to one, 
i.e. $\lVert g\rVert_2=\lVert\gamma\rVert_2=1$.
Furthermore we will not force orthogonality, like
orthogonal transmit signaling ($\langle\gamma_m,\gamma_n\rangle=\delta_{mn}$),
orthogonality at the receiver ($\langle g_m,g_n\rangle=\delta_{mn}$) or 
biorthogonality between transmitter and receiver ($\langle g_m,\gamma_n\rangle\sim\delta_{mn}$).
But note that, advanced equalization techniques (not considered
in this paper) like interference cancellation
will suffer from noise--enhancement and noise correlation 
introduced by non--orthogonal receivers.

\subsection{Complex Schemes}
In this approach full complex data symbols are transmitted according to (\ref{equ:txsignal}).
Depending on the lattice density ($\epsilon<1$) this includes redundancy. 
In the sense of biorthogonality it is then desirable to achieve $\langle g_m,\BH\gamma_n\rangle\sim\delta_{mn}$
for a particular class of channels $\BH$. 
For example the classical OFDM system exploiting a cyclic prefix (cp-OFDM) is obtained by assuming a
lattice generated by $\Lambda=\diag(T,F)$ and setting
$\gamma$ to the rectangular pulse
\begin{equation}
   \gamma(t)=\frac{1}{\sqrt{T_u+T_{cp}}}\Ind{[-T_{cp},T_u]}{t}
   \label{eq:ofdm:gamma}
\end{equation}
The function $\chi_{[-T_{cp},T_u]}$ is the characteristic function of the interval $[-T_{cp},T_u]$, where
$T_u$ denotes the length of the useful part of the signal and $T_{cp}$ the length of the cyclic prefix 
($\approx 10\%T_u$), hence
the OFDM overall symbol period is $T=T_u+T_{cp}$. 
The OFDM subcarrier spacing is $F=1/T_u$.
At the OFDM receiver the rectangular pulse $g(t)=\frac{1}{\sqrt{T_u}}\Ind{[0,T_u]}{t}$ is used which removes
the cyclic prefix. The bandwidth efficiency of this signaling is given as
$\epsilon=(TF)^{-1}=T_u/(T_u+T_{cp})<1$. It can be easily verified that 
$\Amb_{g\gamma}((\tau+m_1T,m_2F))=\sqrt{\epsilon}\delta_{m,0}$ if $0\leq\tau\leq T_{cp}$
(or see \cite{jung:ieeecom:timevariant} for the full formula).
i.e. 
\begin{equation}
   H_{m,n}=\langle g_m,\BH\gamma_n\rangle=\sqrt{\epsilon}\,
   \fourier{h}(m_2F)\delta_{m,n}
   \label{equ:cpofdm:channeldiagonal}
\end{equation}
holds for all channel realization 
as long as the causal scattering function fulfills $B_D=0$ and $\taumax\leq T_{cp}$, 
where $\taumax$ ($B_D$) is its maximal delay (one-sided Doppler) support.
$\fourier{h}(f)$ denotes the Fourier transform of the impulse response 
$h(\tau)=\BHSpread((\tau,0))$
that corresponds to the time-invariant channel. Therefore cp-OFDM is a powerful
signaling, which diagonalizes time-invariant channels, but at the cost of signal power
(the redundancy is not used) and bandwidth efficiency. 

Eq. (\ref{equ:cpofdm:channeldiagonal}) does not hold anymore if the channels 
are doubly-dispersive, as for example modeled by the WSSUS assumptions. 
If considering other pulse shapes independent of a particular realization $\BH$ it is also
not possible achieve a relation similar to (\ref{equ:cpofdm:channeldiagonal}), which will be explained later on.
So it remains to achieve at least $\langle g_m,\gamma_n\rangle\sim\delta_{mn}$ at nearly optimal 
bandwidth efficiency $\epsilon\approx 1$. But one of the deeper results in Gabor theory, 
namely the Balian-Low Theorem (see for example \cite{daubechies:tenlectures}), 
states that (bi-)orthogonal pulses at $\epsilon\approx 1$ must
have bad time-frequency localization properties (at $\epsilon=1$ diverging 
 localization). Indeed,
in discrete implementation the localization of 
orthogonalized Gaussians for $\epsilon\leq1$ and ''tighten'' Gaussians for $\epsilon\geq1$
peaks at the critical density $\epsilon=1$  
so that pulse shaping is mainly prohibited for 
band efficient complex schemes if still (bi-) orthogonality is desired. Nevertheless 
in contrast to cp-OFDM it is via pulse shaping (for $\epsilon<1$) still possible 
to make use of the redundancy.

\subsection{Real Schemes}
\label{subsec:oqam}
For those schemes an inner product $\Real{\langle\cdot,\cdot\rangle}$  is considered, which
is realized by OQAM based modulation for OFDM (also known as OQAM/OFDM) \cite{chang:oqam}. 
It is obtained in
(\ref{equ:txsignal}) and (\ref{equ:receivedsymbol}) with a lattice generated by $\Lambda=\diag(T,F)$
having $|\det\Lambda|=1/2$. 
Before modulation the mapping
$x_n=i^n x^\text{R}_n$ has to be applied \footnote{We 
  use here the notation $i^n=i^{n_1+n_2}$. Furthermore other phase mappings are
  possible, like $i^{n_1+n_2+2n_1n_2}$.},
where $x^\text{R}_n\in\mathbb{R}$ is the real-valued information to transmit.
After demodulation
$\tilde{x}^\text{R}_m=\Real{i^{-m}\tilde{x}_m}$ is performed. Moreover, the pulses $(g,\gamma)$
have to be real. Thus, formally the transmission of the real information
vector $\Bx^\text{R}=(\dots,x^\text{R}_n,\dots)^T$ can be written as
$\tilde{\Bx}^\text{R}=H^\text{R}\Bx^\text{R}+\tilde{\Bn}^\text{R}$
where the real channel matrix elements are:
\begin{equation}
   H^\text{R}_{m,n}=\Real{i^{n-m}H_{m,n}}=\Real{i^{n-m}\langle g_m,\BH \gamma_n\rangle}
   \label{eq:hmn:realschemes}
\end{equation}
and ''real--part'' noise components are $\tilde{n}^\text{R}_m=\Real{i^{-m}\langle g_m,n\rangle}$.
Note that there exists no such relation for OQAM based multicarrier transmission 
equivalent to (\ref{equ:cpofdm:channeldiagonal}) for cp-OFDM. 
Hence, also in time-invariant channels there will be ICI. But in the absence of 
a channel, biorthogonality of the form $\Re\{\langle g_m,\gamma_n\rangle\}=\delta_{m,n}$ 
can be achieved. Furthermore it is known that the design of orthogonal OQAM based multicarrier transmission
is equivalent to the design of orthogonal Wilson bases \cite{boelckei:oqam}. 
Because the system operates with real information at $\epsilon=2$ the effective efficiency is $1$, but
in the view of pulse shaping it is not affected by the Balian--Low theorem.
It is known that the construction of orthogonal Wilson bases is
equivalent to the design of tight frames having redundancy two (which will be explained later on 
in the paper) \cite{daubechies:simplewilson}. 
It will turn out that this equivalence holds also for the
WSSUS pulse shaping problem considered in this paper if  
assuming some additional symmetry for the noise and the spreading function of the channel.
Finally, extensions of classical Wilson bases to non--separable lattices are studied in \cite{kutyniok:generalwilson}.

\section{WSSUS Pulse Design}
In the first part of this section we will collect, what in principle is known in WSSUS pulse shaping theory. 
The result are partially contained in \cite{kozek:thesis,strohmer:lofdm2,jung:ieeecom:timevariant,sayeed:isit2002}.
We begin by establishing a cost function which characterizes the averaged 
performance of uncoded multicarrier transmission
over a whole ensemble of WSSUS channels. 
Even though we consider solely uncoded 
transmission, our results will give insights into the coded performance as well. 
However, an overall optimization of the coded performance is beyond of the
scope of this paper.  
In the aim of maximizing the performance (the cost) we
will show how various design rules on WSSUS pulse shaping occur as steps in this 
optimization problem.
We will explicitely identify the stage at which this argumentation will have a gap which will
be filled by our algorithms presented later on.
In the second part of this section (in \ref{subsec:math:formulation}) 
we will establish a new analytical framework which better 
highlights the underlying algebraic structure. In particular it will turn
out that this is important to understand the appropriate 
optimization strategies presented in the next section.

\subsection{The WSSUS Pulse Design Problem}

In multicarrier transmission most commonly one--tap equalization per
time--frequency slot is considered, hence it is
naturally to require $a$ (the channel gain of the lattice point
$m\in\Indexset$) to be maximal 
and the interference power 
$b$ from all other lattice points to be minimal as possible, where
\begin{equation}
   a\defeq|H_{m,m}|^2\,\,\,\,\text{and}\,\,\,\,
   b\defeq\sum_{n\neq m}|H_{m,n}|^2  
\end{equation}
for complex schemes. For real schemes $H_{m,n}$ has
to be replaced by $H^\text{R}_{m,n}$ from \eqref{eq:hmn:realschemes}.

This addresses the concept of 
{\it pulse shaping}, hence to find good pulses $\{g,\gamma\}$ such that its averaged cross ambiguity
yields maximum channel gain and minimum interference power. 
A comprehensive framework for the optimization of redundant precoders and equalizers with respect to
instantaneous time-invariant channel realizations (assumed to be known at the transmitter) 
is given in \cite{scaglione:redfilt:partA,scaglione:redfilt:partB}.
However in certain scenarios it is much more realistic 
to adapt the pulses only to the second order statistics, given by $\BHScat(\mu)$ 
and not to a particular realization $\BHSpread(\mu)$. Hence, instead
we define the (long term) averaged 
{\it signal-to-interference-and-noise ratio} $\SINR(g,\gamma,\Lambda)$ as
\begin{equation}
   \begin{aligned}
      \SINR(g,\gamma,\Lambda)\defeq\frac{\Ex{\BH}{a}}{\sigmaN+\Ex{\BH}{b}}
   \end{aligned}      
   \label{equ:SINR:def}
\end{equation}
where the averaged channel gain and the averaged interference power are given for
complex schemes as
\begin{equation}
   \begin{split}
      \Ex{\BH}{a}&=\Ex{\BH}{|H_{m,m}|^2}
      =\int_{\Reals^2}|\langle g,\Shift_{\mu}\gamma\rangle|^2 \BHScat(\mu)d\mu\\
      \Ex{\BH}{b}
      &=\sum_{n\in\Indexset\without\{m\}}\Ex{\BH}{|H_{m,n}|^2}
      = \sum_{n\in\Indexset\without\{0\}}\int_{\Reals^2}|\langle g,\Shift_{\Lambda n+\mu}\gamma\rangle|^2 \BHScat(\mu)d\mu
   \end{split}      
   \label{equ:exa:exb}
\end{equation}
Note, that both are independent of $m$. Thus in average all lattice
points have the same $\SINR$. 
Eq. \eqref{equ:SINR:def} will hold also
for the real schemes from Sec.\ref{subsec:oqam} if we assume further that the spreading
function $\BHSpread(\mu)$ for each $\mu$ and the noise process $n(t)$ for each $t$
are circular--symmetric (real and imaginary parts have same variances and are
uncorrelated). We have then for each $m,n$:
\begin{equation}
   \Ex{\BH}{|H^\text{R}_{m,n}|^2}=\frac{1}{2}(\Ex{\BH}{|H_{m,n}|^2+\Real{H_{m,n}^2}})
   =\frac{1}{2}\Ex{\BH}{|H_{m,n}|^2}
   \label{eq:hmn2:realschemes}
\end{equation}
because for circular--symmetric $\BHSpread$ follows $\Ex{\BH}{H_{m,n}^2}=0$. Similarly we
get for each $m$: $\Ex{n}{|\tilde{n}_m^\text{R}|^2}=\tfrac{1}{2}\sigmaN$.
Thus, for complex and real schemes
the optimal time-frequency signaling $\{g,\gamma,\Lambda\}$ 
in terms of $\SINR$ is now given
as the solution of problem
\begin{equation}
   \begin{aligned}
      \max_{(g,\gamma,\Lambda)}{\frac{\Ex{\BH}{a}}{\sigmaN+\Ex{\BH}{b}}}
   \end{aligned}      
   \label{eq:sinroptimization}
\end{equation}
Additional to the norm constraint 
applied on the pulses there has to be a bandwidth efficiency constraint on $\Lambda$ in
the sense of $|\det\Lambda|=\epsilon^{-1}=\text{const}$. 
This stands in contrast to capacity (instead of $\SINR$) optimization with respect to 
scattering knowledge at the transmitter. The capacity optimization problem 
itself is unsolved and it 
is unclear that it could make sense or not to tolerate
slightly increased interference but operate at higher effective rate.

The design problem \eqref{eq:sinroptimization} in this general
constellation is not yet well studied because of its complex structure. Most 
studies in this field are limited to separated optimizations of either $\Ex{\BH}{a}$ or
$\Ex{\BH}{b}$ where the lattice structure $\Lambda$ is assumed to be fixed. 
A comparison between different lattices is given in \cite{strohmer:lofdm2}.
But clearly there must be some connection between  $\Ex{\BH}{a}$ and
$\Ex{\BH}{b}$. For an orthogonal basis this is apparent where the general case can be
established by frame theory \cite{feichtinger:gaborbook,christensen:framesandrieszbases,grochenig:gaborbook}. 
Thus if we define for arbitrary (not necessarily diagonal) $\Lambda$ a Gabor set as 
$\Gabor(\gamma,\Lambda,\Ztwo):=\{\Shift_{\Lambda n}\gamma|n\in\Ztwo\}$ 
we can associate to it a positive semidefinite operator $S_{\gamma,\Lambda}$ as follows
\begin{equation}
   \begin{split}
      (S_{\gamma,\Lambda} f)(t):=\sum_{\lambda\in\Lambda\Ztwo}
      \langle\Shift_\lambda\gamma,f\rangle(\Shift_\lambda\gamma)(t)
   \end{split}      
\end{equation}
If there are constants $A_\gamma>0$ and $B_\gamma<\infty$ such that
for all $f\in\Ltwo(\mathbb{R})$ holds
\begin{equation}
   \begin{split}
      A_\gamma\lVert f\rVert^2_2\leq \langle f,S_{\gamma,\Lambda} f\rangle\leq B_\gamma\lVert f\rVert^2_2
   \end{split}      
   \label{eq:framecondition}
\end{equation}
$\Gabor(\gamma,\Lambda,\Ztwo)$ is said to be a frame
for $\Ltwo(\mathbb{R})$. In this case $S_{\gamma,\Lambda}$ is called the Gabor frame operator associated to 
$\gamma$ and $\Lambda\Ztwo$. If the upper bound in  
(\ref{eq:framecondition}) holds,  $\Gabor(\gamma,\Lambda,\Ztwo)$ is called a Bessel sequence and 
the optimal (minimal) $B_\gamma$ is called its Bessel
bound \cite{christensen:framesandrieszbases}. Clearly $B_\gamma$ is
the operator norm of $S_{\gamma,\Lambda}$ induced by $\lVert\cdot\rVert_2$, 
i.e. $B_\gamma$ is the spectral radius $B_\gamma=\rho(S_{\gamma,\Lambda})$.
If $A_\gamma=B_\gamma$ the frame is called 
{\it tight} and the frame operator is then a scaled identity, i.e. $S_{\gamma,\Lambda}=B_\gamma\Id$. 
In this case and if furthermore $\gamma$ is normalized the Bessel bound represents
the redundancy of
$\{\langle \Shift_{\Lambda n}\gamma,f\rangle|n\in\Ztwo\}$ for a given function $f$.
The redundancy is related to $\Lambda\Ztwo$ only, hence 
tight frames minimize the Bessel bound for fixed $\Lambda$. 
And in this sense tight frames can be seen as
generalization of orthonormal bases for which then would hold $B_\gamma=1$.
For Gabor frames (or Weyl--Heisenberg frames) one has further
\begin{equation}
   1\leq\min_{\lVert\gamma\rVert_2=1} B_\gamma=|\det\Lambda^{-1}|=\epsilon
\end{equation}
For $\epsilon<1$ the set $\Gabor(\gamma,\Lambda,\Ztwo)$ can not establish a frame, i.e. $A_\gamma=0$. But it can
establish a Riesz basis for its span where the minimal Bessel bound is attained for the orthogonal case.
But the latter can be formulated within the frame techniques too.
One important result from Gabor theory is 
the Ron-Shen duality \cite{ronshen:duality} between lattices generated by $\Lambda$ and its adjoint lattice generated with
$\Lambda^\circ\defeq\det(\Lambda)^{-1}\Lambda$. 
The Gabor set $\Gabor(\gamma,\Lambda,\Ztwo)$ establishes
a frame (tight frame) iff the Gabor set $\Gabor(\gamma,\Lambda^\circ,\Ztwo)$ is a Riesz basis 
(ONB\footnote{ONB=Orthonormal basis} basis) for its span. \\
Assuming from now on that $\Gabor(\gamma,\Lambda,\Ztwo)$ is a Bessel sequence
we get immediately from \eqref{eq:framecondition}
\begin{equation}
   \begin{split}
      \Ex{\BH}{a}+\Ex{\BH}{b}&=\sum_{n\in\Indexset}\Ex{\BH}{|H_{0,n}|^2}
      \leq\sum_{n\in\Ztwo}\Ex{\BH}{|H_{0,n}|^2}\\
      &=\Ex{\BH}{\langle g,\BH S_{\gamma,\Lambda}\BH^* g\rangle}
      \overset{\eqref{eq:framecondition}}{\leq} B_\gamma\Ex{\BH}{\lVert\BH^*g\rVert_2^2}
   \end{split}      
\end{equation}
or equivalently $\Ex{\BH}{b}\leq B_\gamma-\Ex{\BH}{a}$ for
$\lVert\BHScat\rVert_1=1$ and $\lVert g\rVert_2=1$ ($\BH^*$ is the adjoint operator of $\BH$
with respect to standard inner product). Similarly follows 
from \eqref{eq:hmn2:realschemes} that
$B_\gamma/2$ is the upper bound for real schemes in the last equation.
This in turn gives for both -- real and complex schemes -- the lower bound 
\begin{equation}
   \begin{aligned}
      \SINR(g,\gamma,\Lambda)=\frac{\Ex{\BH}{a}}{\sigmaN+\Ex{\BH}{b}}\geq\frac{\Ex{\BH}{a}}{\sigmaN+B_\gamma-\Ex{\BH}{a}}
   \end{aligned}      
   \label{eq:sinr:bound}
\end{equation}
used already  in \cite{jung:ieeecom:timevariant}. If $\Gabor(\gamma,\Lambda,\Ztwo)$ is a frame
there is a similar upper bound given with $A_\gamma$. Equality is achieved if $\Indexset=\Ztwo$ and
$\Gabor(\gamma,\Lambda,\Ztwo)$ already establishes a tight frame.
Note that $B_\gamma$ depends on $\gamma$ and 
$\Lambda\Ztwo$ but is independent of the channel where $\Ex{\BH}{a}$ depends on $\{\gamma,g\}$ and
on the channel, but is independent of the lattice again. 
A joint maximization of the lower bound would have similar 
complexity as the original problem, whereas 
separate optimizations of $B_\gamma$ and $\Ex{\BH}{a}$  seems to be much simpler.
Eq. \eqref{eq:sinr:bound} 
motivates a design rule that optimizes the pulse with respect to the channel 
first and performs corrections with respect to a particular lattice afterwards.
Thus we propose the following two-step procedure:
\subsubsection{Step one (Gain optimization)} 
\label{subsec:pulsedesign:gainopt}
In the first step the maximization of the averaged {\it channel gain}
$\Ex{\BH}{a}$ is considered, which is 
\begin{equation}
   \begin{aligned}
      \Ex{\BH}{a}
      &=\int_{\Reals^2}|\langle g, \Shift_{\mu}\gamma\rangle|^2 \BHScat(\mu)d\mu
      =\int_{\Reals^2}|\Amb_{g\gamma}(\mu)|^2 \BHScat(\mu)d\mu
      \leq\lVert\BHScat\rVert_1\lVert g\rVert_2^2\lVert\gamma\rVert^2_2=1
   \end{aligned}      
   \label{eq:channelgain}
\end{equation}
In this context (\ref{eq:channelgain}) was first introduced in \cite{kozek:nofdm1} respectively 
\cite{kozek:thesis}, but similar optimization problems already
occurred in radar literature much earlier. 
In particular for the elliptical symmetry of $\BHScat(\cdot)$ Hermite functions
establish local extremal points as found in \cite{kozek:thesis}. If $\BHScat(\cdot)$ is a
two--dimensional Gaussian the optimum is achieved \emph{only} using Gaussian pulses
for $g$ and $\gamma$  matched in spread and offset to $\BHScat(\cdot)$ (see
\cite{jung:isit06}).
There is a 
close relation to the channel fidelity in quantum information processing which will 
become more clear in Section \ref{subsec:math:formulation}.
In \ref{subsec:gainoptimization:convex:constraint:convex:maximization} 
we will establish the maximization of (\ref{eq:channelgain}) as global-type optimization
problem closely related to bilinear programming. However,
already in \cite{jung:spawc2004} we have proposed the following lower bound 
\begin{equation}
   \begin{split}
      \Ex{\BH}{a}\geq |\langle g,\left(\int_{\Reals^2} \Shift_\mu\BHScat(\mu)d\mu\right)\gamma\rangle|^2
      \defeq|\langle g,\mathcal{L}\gamma\rangle|^2
   \end{split}
   \label{eq:gainlowerbound}
\end{equation}
which admits a simple direct solution given as the maximizing eigenfunctions of
$\mathcal{L}^*\mathcal{L}$ respectively $\mathcal{L}\mathcal{L}^*$.
Furthermore the lower bound is analytically studied in \cite{jung:isit05}.

\noindent{\bf Pulse Scaling:}
The maximization of (\ref{eq:channelgain}) is still a difficult task, numerically and analytically.
However, it is possible to obtain a simple scaling rule by second order approximation of the cross ambiguity function.
For $g$ and $\gamma$ being even and real, the squared cross ambiguity can 
be approximated for small $|\mu|$ as follows (see Appendix \ref{app:crossamb:approximation})
\begin{equation}
   |\Amb_{g\gamma}(\mu)|^2\approx\langle g,\gamma\rangle^2[1-4\pi^2(\mu_2^2\sigma_t^2+\mu_1^2\sigma_f^2)]
   \label{equ:crossamb:approx}
\end{equation}
with 
$\sigma_t^2=\langle t^2g,\gamma\rangle/\langle g,\gamma\rangle$ and 
$\sigma_f^2=\langle f^2\fourier{g},\fourier{\gamma}\rangle/\langle g,\gamma\rangle$.
The latter is a slight
extension of the often used approximation for the auto-ambiguity function ($g=\gamma$), 
which gives ellipses as contour lines of (\ref{equ:crossamb:approx}) 
in the time--frequency plane \cite{wilcox:ambsynthesis}. The optimization problem for the
averaged channel gain turns now into the following scaling problem
\begin{equation}
   \max_{(\gamma,g)}{\int_{\Reals^2} |\Amb_{g\gamma}(\mu)|^2\BHScat(\mu)d\mu}
   \xrightarrow{\langle g,\gamma\rangle=\text{const}}
   \min_{(\sigma_t,\sigma_f)}{\int_{\Reals^2}[\mu_2^2\sigma_t^2+\mu_1^2\sigma_f^2]\BHScat(\mu)d\mu}
\end{equation}
which is an optimization of $\sigma_t$ and $\sigma_f$ only.
For a separable scattering function $\BHScat(\mu)=\BHScat_{t}(\mu_1)\BHScat_{f}(\mu_2)$
this further simplifies to
\begin{equation}
   \begin{aligned}
      \min_{(\sigma_t,\sigma_f)}{\BHScat^{(f)}\sigma_t^2+\BHScat^{(t)}\sigma_f^2}
   \end{aligned}
\end{equation}
where $\BHScat^{(f)}=\lVert\BHScat_t\rVert_1\int\BHScat_f(\nu)\nu^2d\nu$ and 
$\BHScat^{(t)}=\lVert\BHScat_f\rVert_1\int\BHScat_t(\tau)\tau^2d\tau$ are the corresponding scaled 
second moments of the scattering function. 
Optimal points have to fulfill the relation 
\begin{equation}
   \begin{aligned}
      \frac{\partial}{\partial (\sigma_t/\sigma_f)}[\BHScat^{(f)}\sigma_t^2+\BHScat^{(t)}\sigma_f^2]=0
      \Leftrightarrow \frac{\sigma_t}{\sigma_f}=\sqrt{\frac{\BHScat^{(t)}}{\BHScat^{(f)}}}
   \end{aligned}
   \label{equ:pulsescaling:matchingrule:moments}
\end{equation}
Already in \cite{kozek:thesis} this matching rule was found for 
$\BHScat(\mu)=\frac{1}{2B_D\taumax}\Ind{[-\frac{\taumax}{2},\frac{\taumax}{2}]}{\mu_1}\Ind{[-B_D,B_D]}{\mu_2}$ 
(non-causal) and for centered flat elliptic shapes, which can 
easily verified from (\ref{equ:pulsescaling:matchingrule:moments}) to be
$\sigma_t/\sigma_f=\taumax/(2B_D)$.
For this special case the rule was also studied in \cite{liu:orthogonalstf} whereas
the latter derivation of the scaling law in terms of moments needs no further assumptions.
The pulses itself which have to be scaled accordingly were not provided 
by this second order approximation. But in \cite{jung:isit05} 
we have shown that an operator--algebraic formulation  leads to
eigenvalue problem for Hermite-kind differential operator, having 
for small $\BHScat^{(f)}\BHScat^{(t)}$ Gaussians as optimal eigenfunctions.
Moreover this approach gives some hint on the optimal phase space displacement between $g$ and $\gamma$, 
which is also not provided by (\ref{equ:crossamb:approx}).
Hence, what follows is, that Gaussians are a good choice for underspread channels 
($\BHScat^{(f)}\BHScat^{(t)}\ll 1$) if only pulse scaling is considered.

\subsubsection{Step two (Interference minimization)}
\label{subsec:pulsedesign:intmin}

The main objective in this step is the minimization of the upper bound 
on the sum of interferences from other lattice points, i.e. $\Ex{\BH}{b}\leq B_\gamma-\Ex{\BH}{a}$. 
Let us consider a pulse pair $\{g,\gamma\}$ that is returned by step one. Hence they represent
some kind of ''single-pulse channel optimality'' achieved by scaling or direct solution of the gain
optimization problem, i.e. let us say they achieve the value
$F(g,\gamma)\defeq\Ex{\BH}{a}$.
As we will show later on in more detail the optimal value of $F(g,\gamma)$ depends only 
$\gamma$,
where $g$ is given by the ''optimum'', i.e. $F(\gamma)\defeq\max_{\lVert g\rVert_2=1} F(g,\gamma)$ and
$\SINR(\gamma,\Lambda)=\max_{\lVert g\rVert_2=1}\SINR(g,\gamma,\Lambda)$.
Recalling now the Bessel bound $B_{\gamma}=\rho(S_{\gamma,\Lambda})$, we introduce 
a linear transformation $\gamma\rightarrow Q\gamma$ in (\ref{eq:sinroptimization}) 
such that 
\begin{equation}
   \min_{\gamma}\SINR(\gamma,\Lambda)^{-1}\leq\min_{Q,\gamma}\left(\sigmaN+\rho(S_{Q\gamma,\Lambda})\right)/F(Q\gamma)-1
   \label{eq:invsinr}
\end{equation}
To arrive at what is commonly known as pulse orthogonalization, we have to ensure that $F(Q\gamma)\approx F(\gamma)$
which is correct for non-dispersive channels ($F(Q\gamma)=F(\gamma)=1$). 
But for the doubly-dispersive case this could be different and it is exactly the gap which can be
filled by non-orthogonal pulses. Nevertheless, under this assumption we would have 
\begin{equation}
   \min_{\gamma}\SINR(\gamma,\Lambda)^{-1}
   \leq\min_{Q}\left(\sigmaN+\rho(S_{Q\gamma,\Lambda})\right)/\max_\gamma{F(\gamma)}-1
   \label{eq:invsinr}
\end{equation}
where the maximization is solved already by step one. The remaining minimization can be performed 
independently of $\gamma$ as follows.
If $Q=\beta S_{\gamma,\Lambda}^{\alpha}$ and if $\Gabor(\gamma,\Lambda,\Ztwo)$ establishes a frame, we have
\begin{equation}
   \begin{split}
      S_{Q\gamma,\Lambda}
      &=\sum_{\lambda\in\Lambda\Ztwo}\langle\Shift_\lambda Q\gamma,\cdot\rangle\Shift_\lambda Q\gamma
      =QS_{\gamma,\Lambda} Q^*=
      \beta^2 S^{1+2\alpha}_{\gamma,\Lambda}\overset{\alpha=-\frac{1}{2}}{=}\beta^2\Id
   \end{split}
\end{equation}
because $S_{\gamma,\Lambda}$ commute with each $\Shift_\lambda$ for $\lambda\in\Lambda\Ztwo$ so its powers.
Thus, with $\alpha\rightarrow-\frac{1}{2}$ we obtain a tight frame 
which has minimal Bessel bound $\beta^2=|\det\Lambda|$, i.e. we achieved equality in (\ref{eq:invsinr}).
This well known procedure \cite{daubechies:tenlectures} 
was already applied for the pulse shaping problem in \cite{strohmer:lofdm2} and has it origins 
in frame theory. Independently a different method for $\gamma$ being a Gaussian
was proposed in the context of OQAM \cite{lefloch:cofdm} which yields the so called
{\it IOTA pulse} (IOTA=
Isotropic Orthogonal Transform Algorithm).
It is known that IOTA is an equivalent method to obtain a tight frame \cite{janssen:equivalencefab}. 
But note that this method does not work in the general case.
Furthermore because of 
the integer oversampling (two is needed for OQAM) the calculation of 
$S_{\gamma,\Lambda}^{\alpha}$ simplifies much in the Zak-domain
and can be done using efficient FFT-based methods \cite{boelckei:oqam:ortho}. 
The extension to the case where $\Gabor(\gamma,\Lambda,\Ztwo)$ is an incomplete Riesz basis 
is done by Ron-Shen duality. 
In this case the minimal Bessel bound is achieved by an ONB, which is given by 
$\Gabor(S^{-\frac{1}{2}}_{\gamma,\Lambda^\circ}\gamma,\Lambda,\Ztwo)$, i.e.
given by the computation of a tight frame on the adjoint lattice. 
Interestingly the resulting orthogonalization procedure based on duality
is  equivalent to the known Schweinler--Wigner \cite{schweinler:ortho} or 
L\"owdin \cite{lowdin:ortho} orthogonalization.
Hence, we arrive at the following operator 
\begin{equation}
   \begin{split}
      O_{\gamma,\Lambda}=\beta 
      \begin{cases}
         S^{-\frac{1}{2}}_{\gamma,\Lambda}       & |\det\Lambda|\leq 1 \\
         S^{-\frac{1}{2}}_{\gamma,\Lambda^\circ} & \text{else} \\
      \end{cases}
   \end{split}
   \label{eq:orthogonalization}
\end{equation}
to be applied on $\gamma$ to minimize the Bessel bound ($\beta$ ensures the normalization).
To perform this operation a lattice with $\det\Lambda^{-1}=\epsilon$
has to be fixed. In the view of our previous derivations it would be desirable to choose the $\Lambda$
which minimizes
$\delta_1(\Lambda)=|F(O_{\gamma,\Lambda}\gamma)-F(\gamma)|$ which is a rather complicated optimization.
However, it is known that 
$O_{\gamma,\Lambda}\gamma$ is closest to $\gamma$ in the $\ell_2$-sense \cite{janssen:tightwindow}, i.e.
$\min_d{\lVert d-\gamma\rVert_2}=\lVert O_{\gamma,\Lambda}\gamma-\gamma\rVert_2=\delta_2(\Lambda)$.
The relation between $\delta_1(\Lambda)$ and $\delta_2(\Lambda)$ is out of the scope of this paper, but it
is $\delta_1(\Lambda)\rightarrow 0$ whenever $\delta_2(\Lambda)\rightarrow 0$. 

\noindent{\bf Lattice scaling:} 
For $\Lambda=\diag(T,F)$ and $\gamma$ being a Gaussian it is in principle known, that
$T/F=\sigma_t/\sigma_f$ ensures the $\min_\Lambda\delta_1(\Lambda)$.
Moreover, in terms of the channel coherence one can follow the 
argumentation given in \cite{liu:orthogonalstf}, i.e.
\begin{equation}
   \begin{split}
      \sqrt{\BHScat^{(t)}}\leq T \leq \frac{1}{\sqrt{\BHScat^{(f)}}}
      \,\,\text{and}\,\,
      \sqrt{\BHScat^{(f)}}\leq F \leq \frac{1}{\sqrt{\BHScat^{(t)}}}
   \end{split}
\end{equation}
In summary, the overall scaling rule for the lattice and the pulse according to the channel statistics is:
\begin{equation}
   \begin{split}
      T/F=\sigma_t/\sigma_f=\sqrt{\BHScat^{(t)}/\BHScat^{(f)}}
   \end{split}
   \label{eq:lattice:scaling}
\end{equation}

\subsection{Mathematical Formulation} 
\label{subsec:math:formulation}
In the following we will investigate the mathematical structure of the problem more in detail.
Observe that the squared magnitude of the cross ambiguity function
$|\Amb_{g\gamma}(\mu)|^2$ can be written in the following form
\begin{equation}
   \begin{aligned}
      |\Amb_{g\gamma}(\mu)|^2
      &=\langle g,\Shift_{\mu}\gamma\rangle\langle\gamma,\Shift^*_{\mu}g\rangle
      =\Trace{\,G\Shift_{\mu}\Gamma\Shift^*_{\mu}}
   \end{aligned}
\end{equation}
where $G$ and $\Gamma$ are the (rank-one) orthogonal projectors onto $g$ and $\gamma$, i.e.
$Gf=\langle g,f\rangle g$.
The linear functional $\Trace(\cdot)$ denotes the trace, i.e. let us define $\settraceclass$ as the set
of trace class operators. The set
\mbox{$\settraceideal\defeq\{z\,|\,z\in\settraceclass,z\geq0,\Trace{\,z}=1\}$} 
is a convex subset of $\settraceclass$. With $\setrankone$ we will denote the 
\emph{extremal boundary of $\settraceideal$, which is  the set of all orthogonal rank-one projectors}. 
Now let us transform the averaged channel gain furthermore in the following way
\begin{equation}
   \begin{aligned}
      \Ex{\BH}{a}
      =\int_{\Reals^2}\Trace{\{G\Shift_{\mu}\Gamma\Shift_{\mu}^*\}\BHScat(\mu)d\mu}
      =\Trace{\{G\int_{\Reals^2}\Shift_{\mu}\Gamma\Shift_{\mu}^*\BHScat(\mu)d\mu\}}
      \defeq\Trace{\,GA(\Gamma)}\\
   \end{aligned}      
\end{equation}
where we have introduced the map $A(\cdot)$. It maps operators $X$ as follows
\begin{equation}
   \begin{aligned}
      A(X)&\defeq\int_{\Reals^2}\Shift_{\mu}X\Shift_{\mu}^* \BHScat(\mu)d\mu \\
   \end{aligned}
\end{equation}
This integral is mean in the weak sense.
With $K_{\mu}\defeq \sqrt{\BHScat(\mu)}\Shift_{\mu}$ the map $A(\cdot)$ can written
in a standard form known as the Kraus representation $A(X)\defeq\int_{\Reals^2} d\mu K_{\mu}XK_{\mu}^*$,
which establishes a link to {\it completely positive maps}  (CP-maps) \cite{stinespring:positivefunctions}.
CP-maps  like $A(\cdot)$ received much attention due to its application in quantum 
information theory. They represent stochastic maps on the spectrum of $X$.
Let us collect some properties: 
\begin{equation}
   \begin{aligned}
      A\,\text{is unital}\,&\Leftrightarrow A(\Id)=\Id \\
      A\,\text{is trace preserving}\,&\Leftrightarrow \Trace{ A(X)}=\Trace{X} \\
      A\,\text{is hermiticity preserving}\,&\Leftrightarrow A(X^*)=A(X)^*\\
      A\,\text{is entropy increasing}\,&\Leftrightarrow A(X)\minor X
   \end{aligned}
\end{equation}
where $\major$ is in the finite case the partial order due to eigenvalue majorization.
Recall now the gain maximization (step one in \ref{subsec:pulsedesign:gainopt}), i.e.
(\ref{eq:channelgain}) written in the new formulation is
\begin{equation}
   \max_{G,\Gamma\in\setrankone}{\Trace{\,GA(\Gamma)}}\leq 1
   \label{eq:gainoptimization:traceform}
\end{equation}
For the non-dispersive and single-dispersive case 
(all $\sqrt{\BHScat(\mu)}\Shift_\mu$ 
commute pairwise, otherwise we call the channel doubly--dispersive) 
it is straightforward to solve problem (\ref{eq:gainoptimization:traceform}) and it turns out that
the optimal value (the upper bound) is achieved.
This does not hold anymore for the doubly--dispersive case. 
The formulation in (\ref{eq:gainoptimization:traceform}) is 
closely related to the maximization of the quantum channel fidelity criterion in the
area of quantum information theory. In fact - the problems are equivalent if 
considering so called pure states (the rank-one requirement). 
It is quite interesting that the CP-map $A(\cdot)$ considered here corresponds 
for a  Gaussian $\BHScat(\mu)$ to the classical (bosonic) quantum channel (see
\cite{holevo:propquantum,hall:gaussiannoise} and also \cite{arxiv:0409063}). In
\cite{arxiv:0409063,jung:isit06} it is shown that in this case the maximum is achieved 
by so called coherent states 
(time-frequency shifted Gaussians). Furthermore, the important role of Gaussians 
as approximate maximizers is also assessed by the authors in \cite{jung:isit05}. 
The application of cyclic shifts in the trace functional gives
rise to the definition of another map $\tilde{A}(\cdot)$ by
\begin{equation}
   \begin{aligned}
      \Ex{\BH}{a}
      =\Trace{\,\Gamma\int_{\Reals^2}\Shift^*_{\mu}G\Shift_{\mu}\BHScat(\mu)d\mu}
      \defeq\Trace{\,\Gamma\tilde{A}(G)}\\
   \end{aligned}      
\end{equation}
which is the adjoint of $A(\cdot)$ with respect to $\langle X,Y\rangle\defeq\Trace X^*Y$. 
Based on the Weyl--Heisenberg group rules, i.e.
$\Shift^*_\mu        =e^{-i2\pi\mu_1\mu_2}\Shift_{-\mu}\,\,$,
$\Shift_\mu\Shift_\nu=e^{-i2\pi \mu_1\nu_2}\Shift_{\mu+\nu}$ and
$\Shift_\mu\Shift_\nu=e^{-i2\pi(\mu_1\nu_2-\mu_2\nu_1)}\Shift_\nu\Shift_\mu$
we have, that CP-maps with Weyl-Heisenberg structure are covariant with respect to group members, i.e.
\begin{equation}
   A(\Shift_{\mu}\Gamma\Shift^*_{\mu})=\Shift_{\mu}A(\Gamma)\Shift^*_{\mu}
   \label{eq:covariance}
\end{equation} 
Thus, (\ref{eq:gainoptimization:traceform}) is invariant with respect to 
joint time--frequency shifts of $\Gamma$ and $G$. A trivial but important conclusion is that Weyl-Heisenberg (Gabor)
signaling is a reasonable scheme, which guarantees the same averaged
performance on all lattice points.
Another conclusion is that two CP-maps $A_1(\cdot)$ and $A_2(\cdot)$ both having Weyl-Heisenberg structure commute,
i.e. $A_1\circ A_2=A_2\circ A_1$.
Finally, a CP-map $A(\cdot)$ with Weyl--Heisenberg structure 
is self--adjoint ($A=\tilde{A}$) with respect to the inner product 
$\langle X,Y\rangle=\Trace{\,X^*Y}$, whenever $\BHScat(\mu)=\BHScat(-\mu)$.

\noindent Similar transformations can be performed on the interference term, i.e.
\begin{equation}
   \begin{aligned}
      \Ex{\BH}{b}
      &=\Trace{\, G\sum_{\lambda\in\Lambda\Indexset\without\{0\}}\Shift_{\lambda}A(\Gamma)\Shift^*_{\lambda}}
      &\defeq\Trace{(G\cdot(B\circ A)(\Gamma))}
   \end{aligned}      
\end{equation}
The introduced map $B$ 
\begin{equation}
   \begin{aligned}
      B(\Gamma)
      &\defeq \sum_{\lambda\in\Lambda\Indexset\without\{0\}}\Shift_{\lambda}\,\Gamma\,\Shift^*_{\lambda}
      &=\sum_{\lambda\in\Lambda\Indexset}\Shift_{\lambda}\,\Gamma\,\Shift^*_{\lambda}-\Gamma\
   \end{aligned}
\end{equation}
is also hermiticity preserving and fulfills $B(\Gamma)\minor \Gamma$. 
For $|\Indexset|<\infty$ follows $B(\Gamma)\in\settraceclass$ whenever $\Gamma\in\settraceclass$.
Hence, with $\rho\defeq(|\Indexset|-1)^{-1}$ follows $\rho B$ is unital and 
trace preserving. Finally let us define 
\begin{equation}
   \begin{aligned}
      C   \defeq B\circ A\,\,\,\text{and}\,\,\,
      D(X)\defeq\sigmaN+C(X)=C(\sigmaN+X)
   \end{aligned}
\end{equation}
With this definitions the optimization problem reads now
\begin{equation}
   \begin{aligned}
      \max_{X,Y\in\setrankone}{\frac{\Trace{A(X)Y}}{\Trace{D(X)Y}}}
      =\max_{X,Y\in\setrankone}{\frac{\Trace{X\tilde{A}(Y)}}{\Trace{X\tilde{D}(Y)}}} \\
   \end{aligned}  
   \label{eq:optimizationproblem}
\end{equation}

\section{Optimization Strategies and Algorithms}
In this part of the paper we will discuss the desired problems in view of possible optimization strategies
and algorithms. 
Hence we will now consider $\mathbb{C}^L$ as the underlying Hilbert space. The set of trace class operators 
are now represented by $L\times L$ matrices and the set $\settraceideal$ are positive semidefinite matrices having 
normalized trace. The matrix representations of time--frequency shift operators are given as
$\left(\Shift_{\mu}\right)_{mn} =\delta_{m,n\oplus\mu_1}e^{i\frac{2\pi}{L}(\mu_2\cdot m)}$,
where all index--arithmetics are modulo $L$. Thus $\mu\in\setZ_L^2$ where
$\setZ_L:=\{0,\dots,L-1\}$.

\subsection{Convex Constrained Quasi-convex Maximization}
We focus now more in detail on the problem (\ref{eq:optimizationproblem}). 
It is straightforward to see, that one of the optimization variables can be dropped, which is
\begin{equation}
   \begin{aligned}
      \max_{X,Y\in\setrankone}\frac{\Trace{\,A(X)Y}}{\Trace{\,D(X)Y}}
      =\max_{X\in\setrankone}\lambda_{\max}(A(X),D(X))
      =\max_{Y\in\setrankone}\lambda_{\max}(\tilde{A}(Y),\tilde{D}(Y))
   \end{aligned}      
\end{equation}
If we drop $Y$ it is left to maximize the generalized hermitian eigenvalue
$\lambda_{\max}(A(X),D(X))$ or if we drop $X$ it is $\lambda_{\max}(\tilde{A}(Y),\tilde{D}(Y))$.
The maximal generalized hermitian eigenvalue $\lambda_{\max}(A(X),D(X))$ is a quasi convex function in $X$ 
if $D(X)^{-1}$ exists, i.e. all level sets are convex. 
The existence of the inverse is ensured by $\sigmaN\neq0$. Independently it can be shown that for $\sigmaN=0$ this can 
also be achieved with $|\det\Lambda|\leq 1$ and $\Indexset=\Ztwo$. From the quasi convexity follows 
\begin{equation}
   \begin{aligned}
      &\max_{X\in\settraceideal}{\lambda_{\max}(A(X),D(X))}
      =\max_{X_i\in\setrankone}{\lambda_{\max}(A(\sum_ip_iX_i),D(\sum_ip_iX_i))}=\\
      &\max_{X_i\in\setrankone}{\lambda_{\max}(\sum_ip_iA(X_i),\sum_iD(p_iX_i))}
      \leq\max_i\{\max_{X_i\in\setrankone}{\lambda_{\max}(A(X_i),D(X_i))}\}\\
      &=\max_{X\in\setrankone}{\lambda_{\max}(A(X),D(X))}
   \end{aligned}
   \label{eq:generaleigenvalues:convexization}
\end{equation}
Thus, the optimization can be performed over $\settraceideal$ but the set of maximizers contain at least
one $X\in\setrankone$ (the maximum is at least achieved at some vertex). Moreover under our
assumptions ($D(X)^{-1}$ exists) the generalized eigenvalue can be rewritten as the (classical)
eigenvalue maximization problem $\max_{X\in\setrankone}{\lambda_{\max}(A(X)D(X)^{-1})}$.
But note that the argument is now rational in $X$.
However, we aim at maximization of a quasi convex function over a convex set, which is
a global optimization problem (commonly formulated as a minimization problem, i.e. 
quasi-concave minimization). As we will expect that the dimension of the Hilbert space will be large
, i.e. $L\approx 1024\dots8192$, standard techniques for global optimization are mainly 
prohibit. Typically branch--and--bound algorithms are able to find a global optimum for 
low scale problems.
But for the lowest possible dimension $L=2$ the problem can be solved completely
\cite{jung:itgscc05}.
For our setup we will instead propose in the following a simple algorithm which provides a lower bound.
Hence, we make the following mappings explicit
\begin{equation}
   \begin{aligned}
      y(X)&\defeq\arg\max_{Y\in\setrankone}\SINR(X,Y) \\
      x(Y)&\defeq\arg\max_{X\in\setrankone}\SINR(X,Y) \\
      z(X)&\defeq (x\circ y)(X)
   \end{aligned}
\end{equation}
where $\SINR(X,Y)=\Trace{\,A(X)Y}/\Trace{\,D(X)Y}$.
This single--variable maximizations can be efficiently solved by computing the generalized hermitian eigenvalues.
Hence, $y(X)$ (or $x(Y)$) is the generalized maximizing eigenvector of $\{A(X),D(X)\}$ 
(or $\{\tilde{A}(Y),\tilde{D}(Y)\}$).
The mapping $z$ corresponds to one iteration step. The iterative algorithm is given below.
\begin{algorithm}[!]
   \caption{SINR optimization}
   \begin{algorithmic}[1]
      \REQUIRE $\delta>0$
      \REQUIRE an appropriate initialized state $X_0$ (for example a Gaussian)
      \REPEAT
      \STATE Calculate in the $n$'th iteration :
      \begin{equation}
         \begin{aligned}
            X_n &\defeq z^n(X_0)=z(z^{(n-1)}(X_0)) \\
            Y_n &\defeq y(X_n)
         \end{aligned}
      \end{equation}
      \STATE giving functional values $\SINR(X_n,Y_n)=\SINR(X_n,y(X_n))$.\\
      \UNTIL {$\SINR(X_n,Y_n)-\SINR(X_{n-1},Y_{n-1})\leq\delta$}
   \end{algorithmic}
   \label{algo:sinropt}
\end{algorithm}
Convergence in the weak sense is given straightforward by observing that 
$\{\SINR(X_n,Y_n)\}_n$ is monotone increasing and bounded.

\subsection{Convex Constraint Convex Maximization}
\label{subsec:gainoptimization:convex:constraint:convex:maximization}
The following suboptimal strategy to the solution of the problem is very important and related to 
step one in \ref{subsec:pulsedesign:gainopt}. 
It can also be obtained by considering the noise dominated scenario ($\sigmaN\rightarrow\infty$)
\begin{equation}
   \begin{aligned}
      \max_{X\in\setrankone}{\lambda_{\max}(A(X)D(X)^{-1})}
      &=\frac{1}{\sigmaN}\max_{X\in\setrankone}\lambda_{\max}(A(X)(1+\frac{1}{\sigmaN}C(X))^{-1})\\
      &\longrightarrow\frac{1}{\sigmaN}\max_{X\in\setrankone}\lambda_{\max}(A(X))
      \,\,\text{for}\,\,\sigmaN\rightarrow\infty
   \end{aligned}
\end{equation}
As in (\ref{eq:generaleigenvalues:convexization}) there holds the convex relaxation, 
because $\lambda_{\max}(\cdot)$ is a convex function
in its arguments and $A(\cdot)$ is linear. Thus we can instead solve the following convex constraint 
convex maximization problem
\begin{equation}
   \begin{aligned}
      \max_{X\in\settraceideal}{\lambda_{\max}(A(X))} 
      =\max_{Y\in\settraceideal}{\lambda_{\max}(\tilde{A}(Y))} \\
   \end{aligned}      
\end{equation}
Again this type of ''$\max-\max$'' optimization is of global kind, hence methods depend strongly on the
structure of problem. An iterative algorithm but much less computational costly as the iterative maximization of
$\SINR$ is given with 
\begin{equation}
   \begin{aligned}
      y(X)&\defeq\arg\max_{Y\in\setrankone}F(X,Y) \\
      x(Y)&\defeq\arg\max_{X\in\setrankone}F(X,Y) \\
      z(X)&\defeq (x\circ y)(X)
   \end{aligned}
\end{equation}
where $F(X,Y)\defeq\Trace{\,A(X)Y}=\Trace{\,X\tilde{A}(Y)}$.
The maximizations can be solved efficiently by eigenvalues decompositions,
i.e. $y(X)$ and $x(Y)$ are the maximizing eigenvectors of $A(X)$ and $\tilde{A}(Y)$. 
The iterative algorithm is given below.
\begin{algorithm}[!]
   \caption{GAIN optimization}
   \begin{algorithmic}[1]
      \REQUIRE $\delta>0$
      \REQUIRE an appropriate initialized state $X^0$
      \REPEAT
      \STATE Calculate in the $n$'th iteration :
      \begin{equation}
         \begin{aligned}
            X_n &\defeq z^n(X_0)=z(z^{(n-1)}(X_0)) \\
            Y_n &\defeq y(X_n)
         \end{aligned}
      \end{equation}
      \STATE giving functional values $F(X_n,Y_n)=F(X_n,y(X_n))$.
      \UNTIL {$F(X_n,Y_n)-F(X_{n-1},Y_{n-1})\leq\delta$}
   \end{algorithmic}
   \label{algo:gainopt}
\end{algorithm}
The proof of weak convergence is again straightforward.
It has been turned out that this algorithms are extension of the so called ''mountain climbing'' algorithm
proposed by Konno \cite{konno76} for bilinear programming. This can be seen if considering 
a corresponding basis representation. It is known that the set of hermitian matrices establish a real vector space.
Let $\{\sigma_i\}$ be a basis, i.e. we have $X=\sum_i x_i\sigma_i$ and $Y=\sum_j x_j\sigma_j$
\begin{equation}
   \begin{aligned}
      \max_{X,Y\in\setrankone}{\Trace{A(X)Y}}=\max_{x,y\in\setblochrankone}{\langle x,ay\rangle}
   \end{aligned}      
\end{equation}
where the $x=(\dots,x_i,\dots)^T$ and $y=(\dots,y_j,\dots)^T$ and $a$ is a matrix
with elements $a_{ij}=\Trace{\,A(\sigma_i)\sigma_j}$. The optimization problem looks now 
rather simple but the difficulties are hidden in the definition of the set 
$\setblochrankone=\{x\,|\,\sum_i x_i\sigma_i\in\settraceideal\}$
(see Bloch manifolds \cite{arxiv:0502153}). Without going further in detail we can already state, that
if $A(\cdot)$ is self--adjoint ($A\equiv\tilde{A}$, i.e. for $\BHScat(\mu)=\BHScat(-\mu)$)
the matrix $a$ is symmetric. Then the bilinear
programming problem is equivalent to convex quadratic maximization \cite{konno76}
\begin{equation}
   \begin{aligned}
      \max_{x\in\setblochrankone}{\langle x,ax\rangle}
   \end{aligned}      
\end{equation}
Finally let us point out that separate interference minimization is formulated in this framework as
\begin{equation}
   \begin{aligned}
      \min_{X,Y\in\setrankone}{\Trace{\,C(X)Y}}
      =\min_{X\in\setrankone}{\lambda_{\min}(C(X))}
      =\min_{Y\in\setrankone}{\lambda_{\min}(\tilde{C}(Y))}
   \end{aligned}      
\end{equation}
which was already studied in \cite{schafhuber:pimrc02} by means of convex methods. Unfortunately also this problem 
itself is not convex. It is again concave minimization over convex sets, because the convex relaxation
($\setrankone\rightarrow\settraceideal$)
applies here as well. For non--dispersive and single--dispersive channels in turn this can be shown 
to be equivalent to the Bessel bound minimization (\ref{subsec:pulsedesign:intmin}).

\section{Performance Evaluation}
In this section we will evaluate the performance of the proposed pulse shaping algorithms. We compare
them to the performance obtained by the use of properly scaled Gaussians, IOTA function
and rectangular pulses. 
\subsection{WSSUS Grid Matching and Pulse Scaling}
We use a rectangular lattice $\Lambda=\diag(T,F)$ properly scaled to the
WSSUS statistics.
Remember that pulse and lattice scaling with respect to a causal ''flat'' scattering function with
support $[0,\taumax]\times[-B_D,B_D]$ 
in the discrete representation means to fulfill approximately
\begin{equation}
   \frac{\taumax}{2B_D}\approx\frac{\sigma_t}{\sigma_f}\approx\frac{T}{F}\,\,\text{with}\,\,(TF)^{-1}=\epsilon
\end{equation}
according to equation (\ref{eq:lattice:scaling}).
Under a fixed bandwidth constraints $W$ and fixed bandwidth efficiency $\epsilon$ (in complex symbols)
this lattice scaling rule (grid matching) can be easily transformed into an optimal number of subcarriers.
Given $TF=\epsilon^{-1}$, $T/F=\taumax/(2B_D)$ and $F=W/N$, where $N$ is the number of subcarriers,  follows
\begin{equation}
   \begin{aligned}
      N
      &=W\sqrt{\frac{\taumax}{  2\epsilon\cdot B_D}}=W\sqrt{\frac{\taumax c}{2\epsilon \cdot v f_c}}
   \end{aligned}
   \label{equ:optimal:ncarriers}
\end{equation}
where $v$ is the speed between transmitter and receiver, $c$ the speed of light and $f_c$ the carrier
frequency. Moreover, in FFT based polyphase filtering $N$ has be a power of two. 
The rule (\ref{equ:optimal:ncarriers}) represents nothing more than the tradeoff 
between time and frequency division multiplexing in time-variant channels.

\subsection{Performance of the Iterative Algorithms}
We have verified our iterative optimization algorithms for a complex scheme (at $\epsilon=0.5$) 
and a real scheme ($\epsilon=2$ and OQAM)  on pulses of length $L=512$. 
Assuming then a bandwidth $W$ discrete time--frequency shifts $\mu^{(D)}\in\setZ_L^2$ 
are related
to  $\Reals^2\ni\mu=\diag(1/W,W/L)\cdot\mu^{(D)}$, i.e.
\begin{equation}
   \mu^{(D)}_1/\mu^{(D)}_2=\frac{W^2}{L}\cdot \frac{\mu_1}{\mu_2}
   \qquad
   \mu^{(D)}_1\cdot\mu^{(D)}_2=\frac{\mu_1\mu_2}{L}
   \label{eq:finitetranslation}
\end{equation}
If we let $B_D^{(D)}$ and $\tau_D^{(D)}$ be the discrete maximal Doppler shift and
delay spread, the support of the scattering function is then
$[0\dots\tau_d^{(D)})]\times [-B_D^{(D)}\dots B_D^{(D)}]$
of fixed size $P:=(\tau_d^{(D)}+1)(2B_D^{(D)}+1)$.  Only the discrete ratio:
\begin{equation}
   R:=(\tau^{(D)}_d+1)/(2B_D^{(D)}+1)
\end{equation}  
has been varied according to the following table:
\begin{center}
   \begin{tabular}{r|c|c|c|c|c|c|c}
      $\tau_d^{(D)}$ &0  &    1 &    5 &    9 &  29 &  49 &  149 \\\hline
      $B_D^{(D)}$    &74 &   37 &   12 &    7 &   2 &   1 &    0 \\\hline
      $N (\epsilon=0.5)$& 1 &  2 &    8 &   16 &   32 &   64 &  256\\
   \end{tabular}
\end{center}
This gives $P\approx 150$ for all $R$, thus with \eqref{eq:finitetranslation} follows 
$P/L\approx0.29$  which is a rather strong but still underspread channel.
Furthermore the number of subcarriers $N$ is matched to $R$ according to 
\eqref{equ:optimal:ncarriers}
as much as possible, but such that
still $\epsilon=0.5$ and $L\mod N=0$. For OQAM ($\epsilon=2$) the grid matching has been
repeated with the result that the number of subcarriers has to be simply $2N$ to
fulfill the requirements.
Where the minimal and maximal values of $R$ (first and last column in the previous table)
correspond to single-dispersive channels (either time-variant, non-frequency-selective or
time-invariant, frequency-selective) the value in between are fully doubly--dispersive. 
Furthermore a noise power of $\sigmaN=-20$dB is assumed. The $\SINR$--optimal timing-offset 
between $g$ and $\gamma$ for the Gaussians, IOTA and rectangular pulses is verified
to be consistent with our theoretical result \cite{jung:isit05}.
 
\subsubsection{Complex Scheme at $\epsilon=0.5$}
The design criterion here is 
robustness not bandwidth efficiency.
In Fig.\ref{fig:wssus:cmplx:gain:flat} we have shown 
the obtained averaged channel gain $\Ex{\BH}{a}$ for the iterative algorithms in 
comparison to Gaussian pulses, IOTA pulses 
and rectangular pulses. Furthermore the result of lower bound (\ref{eq:gainlowerbound}) is
included, which achieves the same value as the iterative gain optimization.
Remember that IOTA and the tighten version of the iterative result (''gaintight'') represent
orthogonal signaling (via (\ref{eq:orthogonalization})).
The orthogonalization does not really change 
the optimality of the gain optimal solution but significantly reduces the interference as shown
in  Fig.\ref{fig:wssus:cmplx:int:flat}. Finally this increases the $\SINR$ as shown in 
Fig.\ref{fig:wssus:cmplx:sinr:flat}. But the maximal value is achieved directly with
iterative $\SINR$ optimization which yields a {\it non-orthogonal signaling}.

\iffigures
\begin{figure}
   \includegraphics[width=\linewidth]{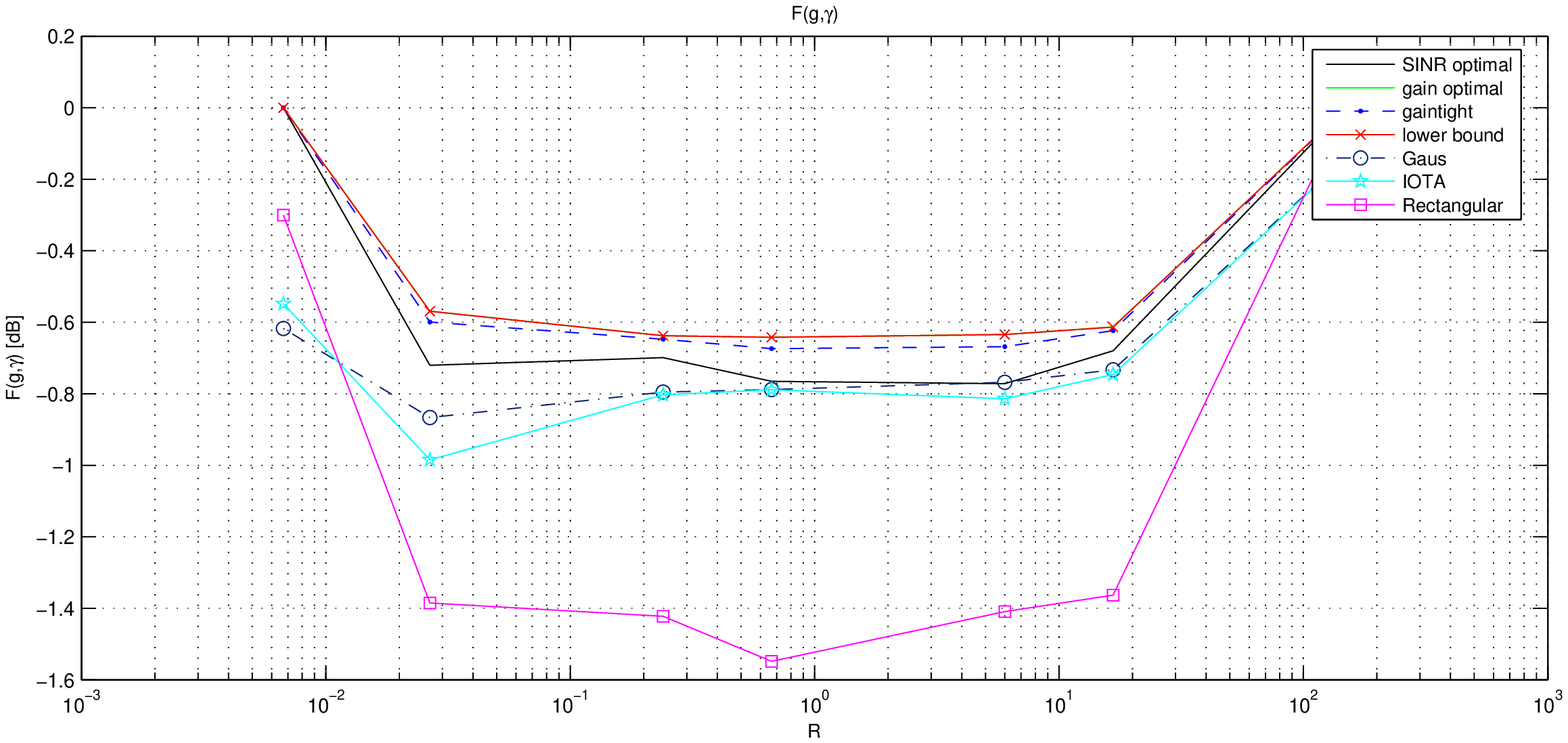}
   \caption{{\it $\Ex{\BH}{a}=F(g,\gamma)$ for a ''flat'' underspread WSSUS channel ($2\tau_d B_D\approx0.29$)
       for a complex scheme (at $\epsilon=0.5$)} - The achieved channel gain is shown for
     the results from the ''SINR optimization'', ''gain optimization'',
     its tighten version via (\ref{eq:orthogonalization}).  The lower bound (\ref{eq:gainlowerbound}) 
     achieves the same value as the pulses obtained from the ''gain optimization''. Furthermore TF-matched Gaussians, 
     TF-matched IOTA and TF-matched rectangular are included. {\it Note that:} ''SINR optimal'' is not ''gain optimal''.
   }
   \label{fig:wssus:cmplx:gain:flat}
\end{figure}
\begin{figure}
   \includegraphics[width=\linewidth]{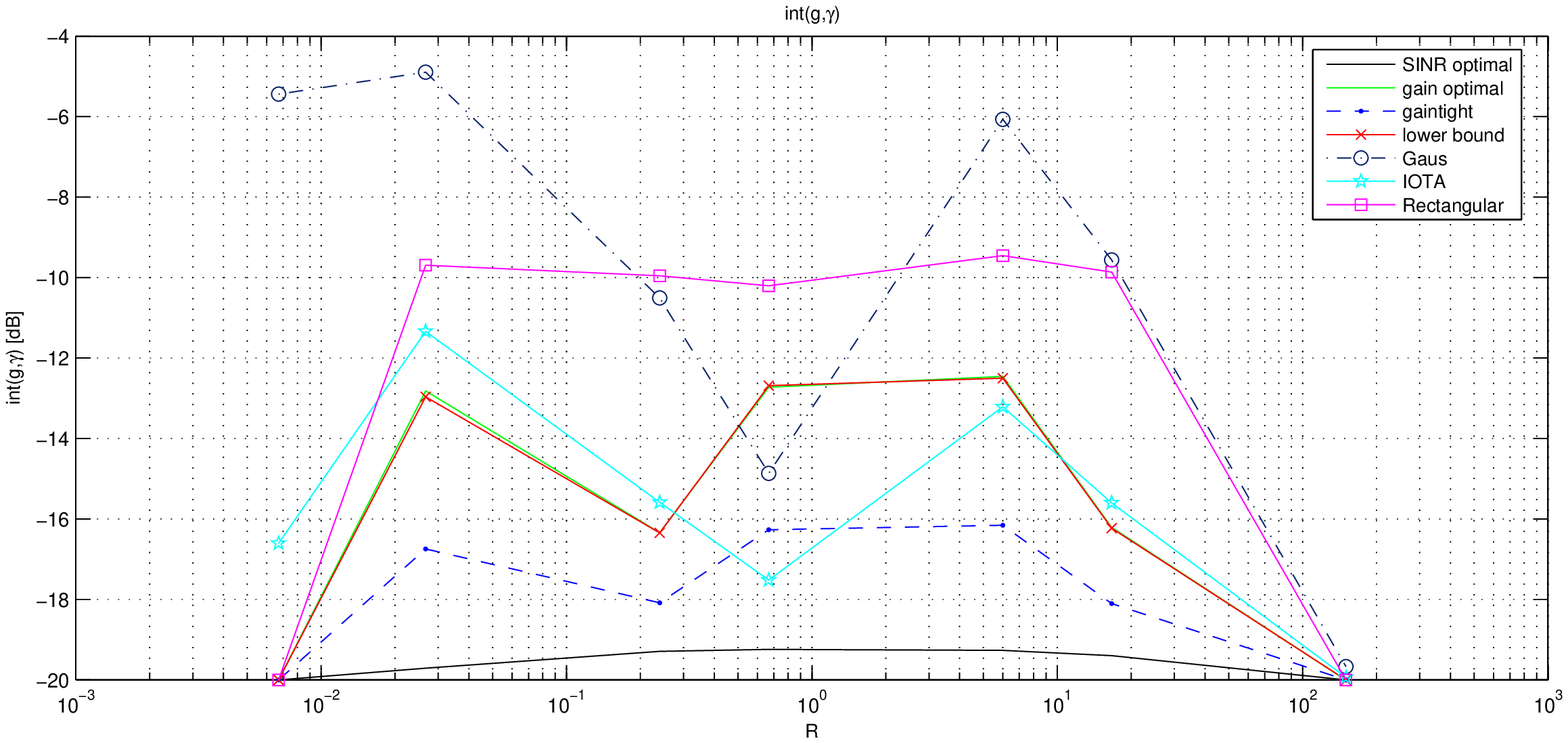}
   \caption{{\it $\Ex{\BH}{b}$ for a ''flat'' underspread WSSUS channel ($2\tau_d B_D\approx0.29$)
       for a complex scheme (at $\epsilon=0.5$)} - The averaged interference power is  shown for 
     the results from the ''SINR optimization'', ''gain optimization'',
     its tighten version via (\ref{eq:orthogonalization}). The lower bound (\ref{eq:gainlowerbound}) 
     achieves the same value as the pulses obtained from the ''gain optimization''. Furthermore TF-matched Gaussians, 
     TF-matched IOTA and TF-matched rectangular pulses are shown. The minimum is achieved with 
     {\it non--orthogonal} pulses.
   }
   \label{fig:wssus:cmplx:int:flat}
\end{figure}

\begin{figure}
   \includegraphics[width=\linewidth]{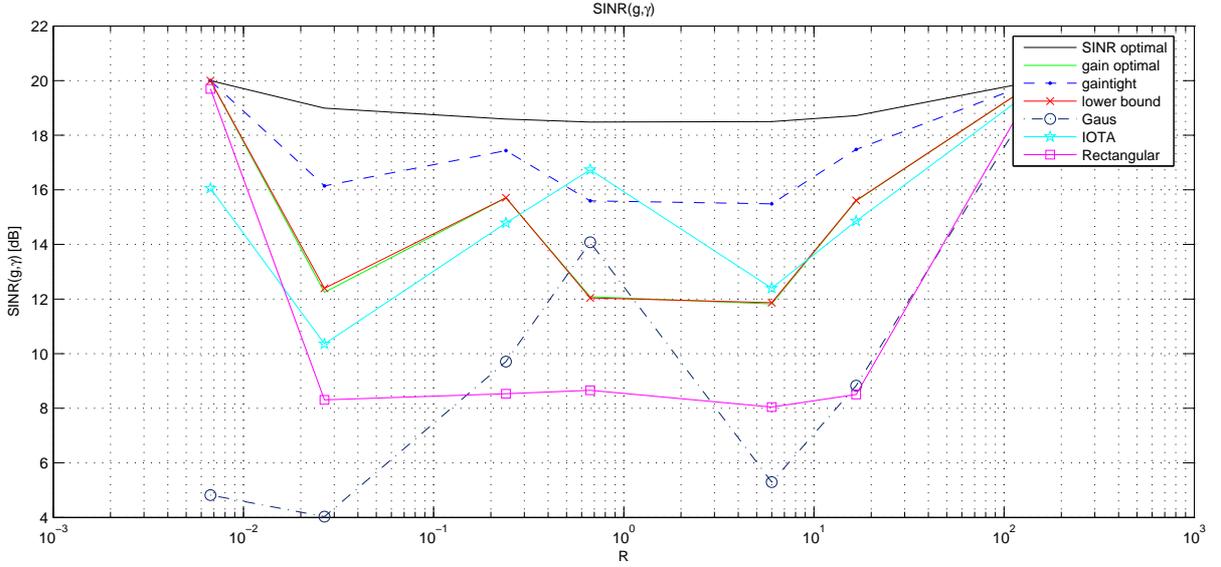}
   \caption{{\it $\SINR(g,\gamma)$ for a ''flat'' underspread WSSUS channel ($2\tau_d B_D\approx0.29$)
       for a complex scheme (at $\epsilon=0.5$)} - The $\SINR$ is shown for
     the results from the ''SINR optimization'', ''gain optimization'',
     its tighten version via (\ref{eq:orthogonalization}). The lower bound (\ref{eq:gainlowerbound}) 
     achieves the same value as the pulses obtained from the ''gain optimization''. Furthermore TF-matched Gaussians, 
     TF-matched IOTA and TF-matched rectangular pulses are shown. The maximum is achieved with 
     {\it non--orthogonal} pulses.
   }
   \label{fig:wssus:cmplx:sinr:flat}
\end{figure}
\fi
\subsubsection{Real Scheme (OQAM)}
The gain optimization does not change for OQAM, hence is the same as for the complex scheme. 
The difference is in the interference term. OQAM operates a maximum spectral efficiency which can
be achieved for linear--independent wave functions with respect to real inner products. Hence there
is no redundancy in the expansion.
From (\ref{eq:sinr:bound}) it is to 
expect that both iterative optimization algorithms must yield similar results, whenever the result
of the gain optimization establishes a nearly tight (snug) frame, i.e. if $B_\gamma/A_\gamma\approx 1$.
This is indeed
the case as shown in Fig.\ref{fig:wssus:oqam:sinr:flat}. The $\SINR$-performance gain with 
respect to the scaling approach 
based on Gaussians (IOTA) is about $3$dB.

\iffigures
\begin{figure}
   \includegraphics[width=\linewidth]{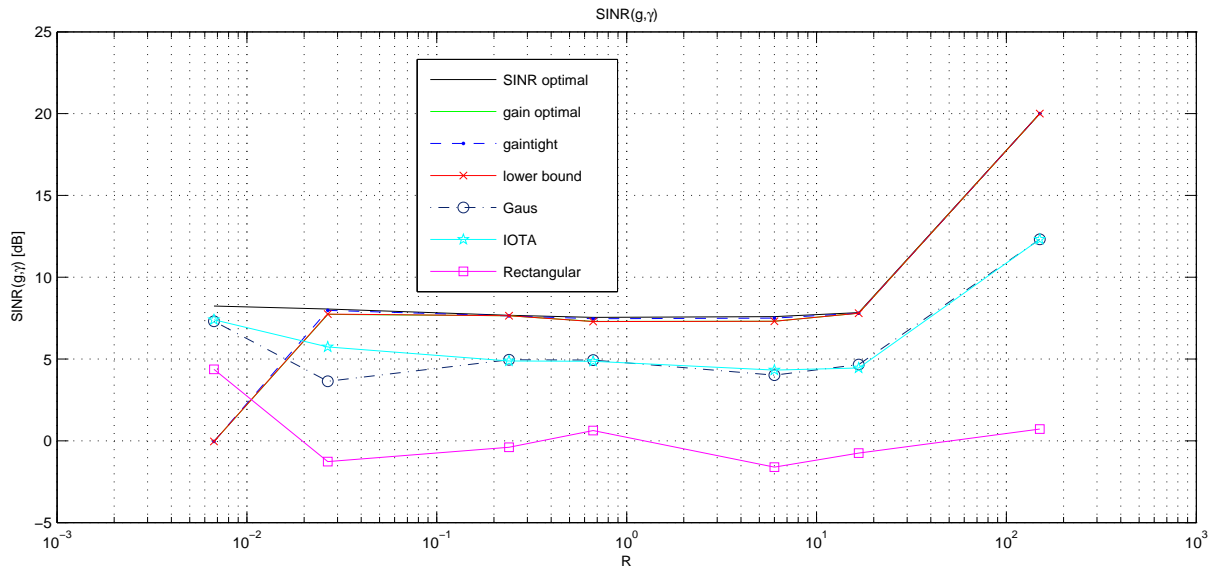}
   \caption{{\it $\SINR(g,\gamma)$ for a ''flat'' underspread WSSUS channel ($2\tau_d B_D\approx0.29$)
       for a real scheme (at $\epsilon=2$ and OQAM)} - The $\SINR$ is shown for
     the results from the ''SINR optimization'', ''gain optimization'',
     its tighten version via (\ref{eq:orthogonalization}). All algorithms and the 
     lower bound (\ref{eq:gainlowerbound}) achieve mainly the same values.
     Furthermore TF-matched Gaussians, 
     TF-matched IOTA and TF-matched rectangular pulses are shown.
   }
   \label{fig:wssus:oqam:sinr:flat}
\end{figure}
\fi

\subsubsection{Implementation Notes}
Note that it is still a numerically challenging task to
apply the iterative pulse design algorithms on real scenarios. 
For that one has to optimize for $L\approx 4N$ and more, where 
the number of subcarriers $N$ has to be according to (\ref{equ:optimal:ncarriers}).
Considering practical scenarios,
for example $\epsilon=2$ (OQAM), $f_c=2$GHz and $W=7.68$MHz, 
it turns out via (\ref{equ:optimal:ncarriers})
that $N=(256,)512,1025,2048(,4096)$ are feasible \cite{jung:inowo2004,jung:iota:techreport}.
Optimizations up to $L=1024$ are in principle possible with conventional direct implementation (using C and {\sc Matlab}
and no FFT processing).
For $L\geq 1024$ only the ''Gain optimization'' and the lower bound remains.
The convergence of the ''Gain optimization'' is rather fast, we achieved the values after $2\dots 3$ iterations.
For the ''$\SINR$ optimization'' which is the most computational expensive task the convergence is slower. We
stopped the algorithm after $5$ iterations.

\section{Conclusions}
We have shown, that pulse shaping with respect to the second order statistics of a WSSUS channel
is challenging optimization problem. We have introduced a new theoretical framework which straightforwardly 
yields two abstract optimization problems which can be partially related to other areas
of quantum physics and mathematics. Unfortunately, due to non--convexity of these 
large scale optimization problems
no standard methods applies.
In fact, global solutions will strongly rely on the structure and 
can be obtained only for special cases.
But with the presented iterative algorithms we have verified that even for
an advanced multicarrier transmission like OQAM/IOTA potential improvement ($3 - 6$dB in $\SINR$
for strong doubly--dispersive channels) can be expected. 
Moreover, it is very likely that additional gain can be obtained if also 
advanced receiver structures instead of one--tap equalizers
are used. Those will profit much more from the sparsity of the effective channel decomposition. 


\appendix

\subsection{The $\order(2)$-approximation for  cross ambiguity functions}
\label{app:crossamb:approximation}
In this part we will provide a slight variation of a well known approximation for auto ambiguity
functions, in the context of radar theory probably first time presented in \cite{wilcox:ambsynthesis}. 
Recall that the cross ambiguity function of $g$ and $\gamma$ is given as
\begin{equation}
   \Amb_{g\gamma}(\mu)=\langle g,\Shift_{\mu}\gamma\rangle
   =\int\overline{g(t)}e^{i2\pi\mu_2 t}\gamma(t-\mu_1)dt
\end{equation}
If we plug in the following series expansions
\begin{equation}
   \begin{split}
      \gamma(t-\mu_1)=\gamma(t)-\mu_1\dot{\gamma}(t)+\frac{1}{2}\mu_1^2\ddot{\gamma}(t)+\order(2)\,\,\text{and}\,\,\,\,
      e^{i2\pi\mu_2 t}=1+i2\pi\mu_2 t-2\pi^2\mu_2^2t^2+\order(2)
   \end{split}
\end{equation}
we get the following approximation for the cross ambiguity function
\begin{equation}
   \begin{split}
      \Amb_{g\gamma}(\mu)=
      &
      \langle               g,\gamma\rangle+
      i2\pi\left( \mu_2  \langle g,t  \gamma\rangle+\mu_1  \langle\fourier{g},f  \fourier{\gamma}\rangle\right)+\\
      &
      2\pi^2\left(\mu_2^2\langle g,t^2\gamma\rangle+\mu_1^2\langle\fourier{g},f^2\fourier{\gamma}\rangle\right)+
      i2\pi\mu_1\mu_2\langle g,t\dot{\gamma}\rangle+\order(2)
   \end{split}
\end{equation}
where $t$, $f$ are multiplication operators (with the variables $t$ in the time domain and with $f$ in the Fourier domain).
The functions $\fourier{\gamma}$ and $\fourier{g}$ denote the Fourier transforms of $\gamma$ and $g$.
Let us furthermore assume that $g\cdot\gamma$ and $\fourier{g}\cdot\fourier{\gamma}$ are symmetric. This is fulfilled
if for example $g$ and $\gamma$ are real and itself symmetric. Then we have
\begin{equation}
   \begin{split}
      \Amb_{g\gamma}(\mu)=
      \langle g,\gamma\rangle\cdot\left(1-2\pi^2(\mu_2^2\sigma_t^2+\mu_1^2\sigma_f^2)+
        i2\pi\mu_1\mu_2\frac{\langle g,t\dot{\gamma}\rangle}{\langle g,\gamma\rangle}\right)+\order(2)
   \end{split}
\end{equation}
where $\sigma_t^2=\langle t^2g,\gamma\rangle/\langle g,\gamma\rangle$ and 
$\sigma_f^2=\langle f^2\fourier{g},\fourier{\gamma}\rangle/\langle g,\gamma\rangle$. For
the squared magnitude of $\Amb_{g\gamma}$ we find
\begin{equation}
   \begin{split}
      |\Amb_{g\gamma}(\mu)|^2=
      |\langle g,\gamma\rangle|^2\cdot\left(1-2\Re{\left(2\pi^2(\mu_2^2\sigma_t^2+\mu_1^2\sigma_f^2)-
          i2\pi\mu_1\mu_2\frac{\langle g,t\dot{\gamma}\rangle}{\langle g,\gamma\rangle}\right)}\right)+\order(2)
   \end{split}
\end{equation}
Finally let us force $g$ and $\gamma$ to be real, which gives the desired result
\begin{equation}
   \begin{split}
      |\Amb_{g\gamma}(\mu)|^2=
      \langle g,\gamma\rangle^2\cdot\left(1-4\pi^2(\mu_2^2\sigma_t^2+\mu_1^2\sigma_f^2)\right)+\order(2)
   \end{split}
\end{equation}

\bibliographystyle{IEEEtran}
\bibliography{references}

\end{document}